\documentclass[11pt]{article}
\usepackage{geometry}                
\usepackage{graphicx}
\usepackage{amssymb}
\usepackage{epstopdf}
\usepackage{color}

\title{Brief Article}
\author{The Author}

\begin{document}
\noindent
{\bf Men will stretch out their eyes:\\ Or, What happened to Christopher Wren's inaugural\\[10pt]\rm Robert I McLachlan\footnote{School of Fundamental Sciences, Massey University, New Zealand. {\tt r.mclachlan@massey.ac.nz}\\Published in {\em The Bulletin, Society for the History of Astronomy} Issue 29 pp. 39--42, 2018.}\\[10mm]

\parskip=6pt
I often find myself in need of  a snappy quote about the future. Several times I've used
the famous
``prediction is difficult, especially about the future\footnote{used by Neils Bohr,
although its origin is unknown; see  {\tt http://quoteinvestigator.com/2013/10/20/no-predict}}'', and that one still makes me chuckle when
I say it to myself. This time I wanted something more related to astronomy, and I was stunned
and thrilled to find the following line attributed to Christopher Wren:

\bigskip
{\narrower\noindent\em A time will come when men will stretch out their eyes. They should see planets like our Earth.

}
\bigskip

It even came with a reference: Inaugural lecture, Gresham College, 1657. At a time when, after a long
wait, we really are seeing planets like our Earth, when thousands of extrasolar planets have been detected%
\footnote{3440 as of January 2017; see {\tt http://exoplanetarchive.ipac.caltech.edu}},
some in systems of up to seven planets; when  new land- and space-based extrasolar planetary
observatories are opening every year; when extrasolar atmospheric spectroscopy is already a reality with
the best yet to come, including a chance at the detection of widespread complex extrasolar life%
\footnote{Jeff Hecht, The truth about exoplanets, {\em Nature} vol. 530 pp. 272--274, 17 Feb 2016}
,---this voice from the dawn of the telescope, 360 years ago, seemed strikingly prescient.

All I knew about Christopher Wren was that he rebuilt St Paul's Cathedral and numerous London churches after the great fire of London. But that was in 1666, nine years after the alleged quote. My curiosity was piqued.

With such a specific reference, to such a famous person, subject of a stream of books and articles running continuously from his lifetime to the present, I thought it would only take a few minutes (seconds, even) to track down. It turned out to be months. And although, to give away the ending, the quote turned out to be partly a dud, the search itself was marvellously rewarding.

It turns out that Wren's inaugural lecture is extremely well known to those who know about Wren or about the history of astronomy. It made his name in science and led to an Oxford professorship. Only after 1665 was he gradually
drawn into architecture and public works. This might not have happened: his son wrote

{\narrower\noindent\em Hence it was, that in the latter part of his life, he has been often heard to complain; that King Charles the 2nd: had done him a disservice in taking him from the pursuit of those studies, and obliging him to spend all his time in rubbish; (the expression he had for building:) for, had he been permitted to have follow'd the Profession of Physick, in all probability he might have provided much better for his family.\footnote{Parentalia}

}

After seven years as a student and fellow in Oxford, Wren was appointed Professor of Astronomy at Gresham College, London, on 7 August 1657, aged just 24. Gresham College was founded in 1597 to provide free lectures to the public, a role it still performs today. His lecture was delivered in Latin, but, fortunately for me, he drafted it first in English and this version was published in the rather gloriously titled {\em Parentalia: or, Memoirs of the Family of the Wrens; Viz. of Mathew Bishop of Ely, Christopher Dean of Windsor, \&c. but chiefly of Sir Christopher Wren, Late Surveyor-General of the Royal Buildings, President of the Royal Society, \&c. \&c.}\footnote{subtitled {\em In which is contained, besides his Works, a great Number of Original Papers and Records; on Religion, Politicks, Anatomy, Mathematicks, Architecture, Antiquities; and most Branches of Polite Literature}, London 1750. The draft inaugural lecture appears on pages 200--206.}

I checked the English version of the inaugural lecture via Google Books. I didn't see the quote. The first half of the quote was there, in altered form (``a Time would come, when Men should be able to stretch out their Eyes\dots''); but no ``planets like our Earth''.

I checked the Latin version\footnote{published in John Ward, {\em The Lives of the  Professors of Gresham College}, London, 1740, App. 8 pp. 29--39; Google Translate made a complete hash of this, but  a friendly Latin teacher pointed me to a translation at {\tt  http://www.philological.bham.ac.uk/wren2/trans.html}}.
 It's quite different from the English draft, although it follows the same outline. Neither the ``planets like our Earth'' nor the ``stretch our their eyes'' parts were present in any form.
 
This was getting odd.

Then I found a recent Gresham Lecture by Anthony Geraghty, {\em From astronomy to architecture: Sir Christopher Wren, Gresham College, and the rebuilding of the city churches after the great fire of London}, delivered
on 30 June 2015. The lecture discusses Wren's inaugural in great detail. There is a similarly detailed discussion in
Adrian Tinniswood's 2010
{\em His Invention So Fertile: a Life of Christopher Wren}:

{\em \narrower\noindent
We must constantly remind ourselves that it comes from  a young man of only twenty-four, so learned, witty, confident and passionate is it, showing both a scholarly understanding of recent developments in astronomy and the sheer, naked joy Wren felt at the new worlds that were opening up as a result.

}

Clearly the lecture was something important. But where was the quote?

I did find the quote, and it is (sort of) in the lecture. I first found the critical passage in an article by Michael Hunter,
well worth reading in its entirety, {\em The Making of Christopher Wren}\footnote{%
The London Journal, vol. 16, no. 2, pp. 101--116, 1991.}. This article also discusses the lecture in detail, but, importantly for me, in does include the crucial phrase ``planets like our Earth.'' To get the context and the glory of it, here is Wren's entire passage. Cast your mind back to 1657:

{\em\narrower\noindent
I cannot (most worthy Auditors) but very much please myself in introducing Seneca, in his Prophecy of the new World, Ñ

{\narrower\noindent Ages will come in future years,\\ in which the ocean will loosen nature's bonds, \\the great earth will be flung open, \\and Tiphys will reveal new worlds, \\nor will Thule be the world's end.

}

But then I only begin to value the Advantages of this Age in Learning 
before the former, when I fancy him continuing his Prophecy, \& imagine 
how much the ancient laborious Enquirers would envy us, should he 
have sung to them, that {\color{red}a Time would come, when Men should be able 
to stretch out their Eyes} as Snails do, \& extend them to fifty feet in length ; 
by which means, they should be able to discover Two thousand Times 
as many Stars as we can ; and find the Galaxy to be Myriads of them; and 
every nebulous Star appearing as if it were the Firmament of some other 
World, at an incomprehensible Distance, bury'd in the vast Abyss of 
intermundious Vacuum: That {\color{red} they should see the Planets like our Earth},
unequally spotted with Hills and Vales: that they should see Saturn, a very Proteus, 
changing more admirably than our Moon, by the various Turnings, and 
Inumbrations of his several Bodies, \& accompany'd besides with a Moon 
of his own ; that they should find Jupiter to be an oval Earth, whose Night 
is enlighten'd by four several Moons, moving in various Swiftnesses, and 
making Multitudes of Eclipses: That they should see Mars, Venus, and 
Mercury to wax and wain: And of the Moon herself, that they should 
have a Prospect, as if they were hard by, discovering the Heighths and 
Shape of the Mountains, and Depths of round and uniform Vallies, the 
Shadows of the Mountains, the Figure of the Shores, describing Pictures 
of her, with more Accurateness, than we can our own Globe, and therein 
requiting the Moon for her own Labours, who to discover our Longitudes, 
by eclipsing the Sun, hath painted out the Countries upon our 
Globe, with the point of her conical Shadow, as with a Pencil. After all 
this, if we should have told them, how the very fountain of Light is variegated 
with its Faculae and Maculae, proceeding round in regular Motions, would 
not any of the Astronomers of his Time have chang'd their 
whole Life for a few windy Days, (in which principally the Solar spots 
appear) or a few clear Nights of our Saeculum.

}

Only now did I begin to see what had happened. First, the passage ``Planets like our Earth''
really does appear. Why did I miss it? It turns out that I had been looking at the wrong
source. The lecture was also reprinted in the first biography of Wren, {\em Memoirs
of the Life and Works of Sir Christopher Wren}, by James Elmes, published in 1823.
For reasons I can't fathom, unless it was a simple mistake, the inaugural lecture printed
 there is complete except for one single phrase. The words {\em ``That they should see the Planets like our Earth,
unequally spotted with Hills and Vales:''}, the ones I was becoming obsessed by, were dropped.
They were also dropped in Elmes's second work on Wren, {\em Sir Christopher Wren and his times}, London 1852, and in a 1903 edition. (The book seems pretty popular. It was reprinted in 2015 by Cambridge University Press, presumably with the same error.)

OK. Now I had my source. But compare the original and the modern versions. The modern
quote is not only abridged, {\em it completely reverses the sense of the original}. Wren is not
remotely speculating about future extrasolar planets, or even about the future at all.
He is talking about how remarkable the astronomy {\em of the year 1657} would seem to the
ancients, if they had but known about it. He is singing the praises of the present (and well he might). 
The crucial edits are changing {\em A time would come}
to {\em A time will come} and changing {\em the planets} (i.e., the planets of our solar system)
to just {\em planets}.

The brilliant ``stretch out their eyes'' phrase, which made me think about humankind throwing our
vision right across the visible universe, is also diminished in the original (though still a great phrase).
{\em Men should be able 
to stretch out their Eyes as Snails do, \& extend them to fifty feet in length}; the very next year, in 1658, Wren did in fact erect a 35 foot telescope in the grounds of Gresham College.\footnote{%
This telescope was given to the Royal Society by Sir Paul Neale.
A full account of Wren's astronomical work, and the events that led to his appointment at Gresham,
can be found in J A Bennett, Christopher Wren: Astronomy, Architecture, and the Mathematical Sciences,
{\em Journal for the History of Astronomy \bf 6} (1975), p.149--184.
Huygens visited Wren and other English astronomers in 1661 and used this telescope to observe the
moon. ``But this telescope of 35 feet did not seem to me as distinct as mine of 22, of which I promised to bring the glass.'' ({\em Oeuvres Compl\`ete de Christiaan Huygens}, vol. 15, Nijhoff 1888, p. 70; {\tt 
 https://archive.org/stream/oeuvrescompltesd15huyg\#page/70/mode/2up}).
Even more to the point, Wren wrote in {\em De Corpore Saturni} c. 1658, ``it was granted to us to have the use of very well worked telescopes
of 6, 12, 22 and even 35 feet long\dots [Sir Paul Neile] is the man who, having hired the best workmen,
ordered the making of these above mentioned celestial device, and even greater ones, of 50 feet, 
in his own house.'' The 50 foot telescope may not have been very successful: 
Robert Hooke wrote in a history of telescopes, ``Sir Paul Neile made some [lenses] of 36 Foot pretty good, and one of 50, as I have been informed, but not answerable.''}

The hunt had now become a whodunnit. Who found, who abridged, and who edited the quote? The fake quote has by now multiplied out of all proportion, appearing all over the internet, in numerous textbooks, astronomy books, and books of quotations, probably for the same reason that I was so taken with it myself in the first place.

As far as I can determine, the actual fake quote first appeared in two books, both published in 2002:
{\em The Expanding Universe: A Beginner's Guide to the Big Bang and beyond}, by Mark Garlick (DP Publishers, April 1 2002) (page 48), and in an identical form in {\em If the Universe Is Teeming with Aliens, Where Is Everybody?: 50 Solutions to Fermi's Paradox \& the Problem of Extraterrestrial Life} by Stephen Webb (Copernicus, October 4 2002, page 150). In both cases, the quote illustrates discussions of extrasolar planets. After that, it starts to appear all over. 

But that wasn't the real origin. Chapter 5 of {\em Cosmos}, by Carl Sagan---the 1980 book that accompanied the enormously popular and influential TV series, watch by as many as 400 million people---opens with 3 epigraphs.
The first is from the Sumerian epic {\em Enuma Elish}, c. 2500 BC: {\em In the orchards of the gods, he watches the canals\dots}. The second is the opening paragraph of Christian Huygen's {\em New Conjectures Concerning the Planetary Worlds, Their Inhabitants and Productions} from 1690 (after Wren's speech, but during Wren's lifetime); and the third is

{\narrower\em\noindent
A time would come when Men should be able to stretch out their Eyes\dots they should see the Planets like our Earth.

}

So I think the story starts with Carl Sagan. He gets the quote right, including the ellipses. He gets the context right, linking it to Huygen's speculations on the solar system, and placing it in a chapter on Mars; perhaps I could query its close association with the Huygens: Wren was only talking about the planets being {\em ``spotted with Hills and Vales''}, not  inhabited. In addition, {\em Cosmos} is incredibly erudite, every chapter being studded with classical references and historical context. In some ways Sagan's literary style seems closer to Wren's than to what would be produced in a similar book today. It's entirely possible to believe the Sagan found Wren's quote himself, either in the original source or in a history of astronomy. And his book in turn proved so popular that either Garlick or Webb (or some earlier author I haven't noticed) could easily have seen it there and---perhaps unconsciously---adapted it.

But I want to end with Wren, and to encourage you to read his only surviving lecture on astronomy.\footnote{%
The (correct!) lecture can be found in {\em Parentalia}, {\tt https://books.google.co.nz/books?id=Tm1MAAAAcAAJ}, pp. 200--206.}
There is just so much in it, and so much of the intellectual ferment of the 17th century. There is a lot just in the passage quoted above. In fact, consider just one line,

{\narrower\em\noindent
every nebulous Star appearing as if it were the Firmament of some other 
World, at an incomprehensible Distance, bury'd in the vast Abyss of 
intermundious Vacuum\dots\footnote{Here {\em intermundious} is Wren's translation of the Latin {\em intermundiis}. The normal anglicization is intermundane, meaning between worlds. The Latin version of the inaugural has {\em some cloud of stars, this would rather be a firmament, not, perhaps, ours, but that of some very remote universe separated by vast interstellar distances\dots.}}

}

The `nebulous stars' were known to Ptolemy, and had been resolved into stellar clusters by Galileo. By but Wren's time two other objects had been discovered that could not be resolved, the great nebulae in Andromeda and in Orion. These may have prompted Wren's speculation that the nebulae are other galaxies (``firmaments''). 
Gerald Whitrow calls this ``the earliest suggestion\dots of such a hypothesis relating to so-called nebulous stars''.\footnote{G J Whitrow, Kant and the Extragalactic Nebulae, Quarterly Journal of the Royal Astronomical Society, Vol. 8, pp. 48--56, 1967. Kant made  detailed speculations along these lines in 1755---having
worked on the tidal slowing of the Earth's rotation the previous year---before moving into other areas. The extragalactic nature of the nebulae was not confirmed until Edwin Hubble measured the distance to Andromeda in 1923.}

{\em Imagine 
how much the ancient laborious Enquirers would envy us\dots}

\newpage

\begin{center}
\includegraphics[width=0.6\textwidth]{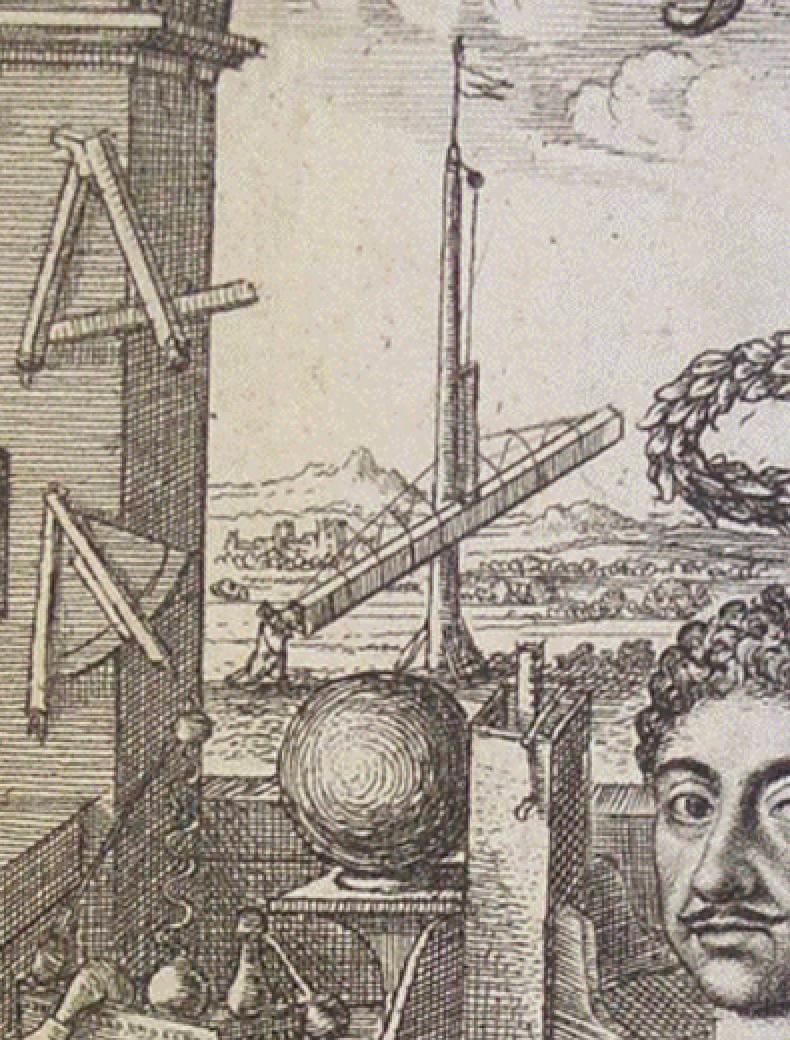}\\[5mm]

\begin{minipage}{0.6\textwidth}
Robert Hooke's 60 foot telescope of 1666, from the frontispiece to Thomas Sprat's 1667 {\em History of the Royal Society of London}, engraving designed by John Evelyn; this telescope is hanging from the mast erected
in 1658 for the 35 foot telescope. This fascinating engraving is so significant  that it is the subject
of an entire book, {\em The Image of Restoration Science} by Michael Hunter, Routledge 2016.
\end{minipage}
\end{center}

\end{document}